# Magnetic / dielectric anomalies and magnetodielectric / magnetoelectric effects in Z- and W-type hexaferrites

J. Li,[1] H.-F. Zhang,[1] G.-Q. Shao,[1,a)] W. Cai,[1] D. Chen,[1] G.-G. Zhao,[1] B.-L. Wu,[2] and S.-X. Ouyang[3]

[1]*State Key Laboratory of Advanced Technology for Materials Synthesis and Processing, Wuhan University of Technology, Wuhan, 430070, China*

[2]*Key Laboratory of New Processing Technology for Nonferrous Metals and Materials, Guilin University of Technology, Guilin, 541004, China*

[3]*China Building Materials Academy, Beijing, 100024, China*

Two kinds of specimens, with the major phase of $Sr_3Co_2Fe_{24}O_{41}$ ($Sr_3Co_2Z$) and $SrCo_2Fe_{16}O_{27}$ ($SrCo_2W$) hexaferrites respectively, were fabricated through solid-state reaction. The phase composition, magnetic and dielectric properties, magnetodielectric (MD) effect, magnetoelectric (ME) effect and pyroelectric properties were studied. Results show that magnetic and dielectric anomalies are induced by the magnetocrystalline anisotropy (MCA) transition. They can be considered as characteristic properties (*e.g.*, $Sr_3Co_2Z$ at 370 K) but are not a sufficient condition for the MD and ME coupling. The T-block structure, existing in $Sr_3Co_2Z$ but absent in $SrCo_2W$, results in the dielectric response with ferroelectric (FE) and magnetic contributions.

## I. INTRODUCTION

$Sr_3Co_2Z$, belongs to Z-type hexaferrite, shows ME (magnetically induced ferroelectricity) and MD (magnetically induced dielectric constant change) effects at a low magnetic field ($H$) and room temperature (RT).[1-3] It has attracted much attention due to the potential technological applications. The hexaferrites are classified into six types (M, Y, Z, W, X and U) according to different stacking sequences of three fundamental blocks (S-, R- and T-blocks) in their crystal structure.[1,2,4,5] The Z-type hexaferrite is formed according the RSTSR*S*T*S* sequence (the asterisk indicates those 180° rotated around *c*-axis) and W-type according the RSSR*S*S* sequence. Investigations on $Sr_3Co_2Z$ have been done in some aspects.[1,3,6-10] However, the early studies on their dielectric properties were mainly focused on high frequencies (> 10 MHz) at RT for applications in microwave field.[11-14] Only two papers involved $\varepsilon'(T)$ (temperature dependence of dielectricity) at low frequencies and / or in a magnetic field.[1,3] Besides, all the hexaferrites with MD / ME coupling effects are found to be on the M–Y line in (Ba, Sr)O–$Fe_2O_3$–$Me$O ternary diagram (Y-, Z- and U-type, see Fig. 1, *Me* denotes the divalent metal ion).[15,16] Due to the synthesis difficulties,[1,6,17,18] the reported samples often co-exist with

---

a)Author to whom correspondence should be addressed. Electronic mail: gqshao@whut.edu.cn.

impurities on the M–S line (W- and / or X-type, Fig. 1).[15,16] So it is a meaningful work to investigate the quantity-known samples which contain simultaneously the compounds on both lines. Target materials in this work are specimens with the major phase of Z- and W-type hexaferrites, respectively. The phase composition, magnetic and dielectric properties, MD effect, ME effect and pyroelectric properties were studied. Correlation between magnetic and dielectric anomalies was investigated considering their crystal and magnetic structures.

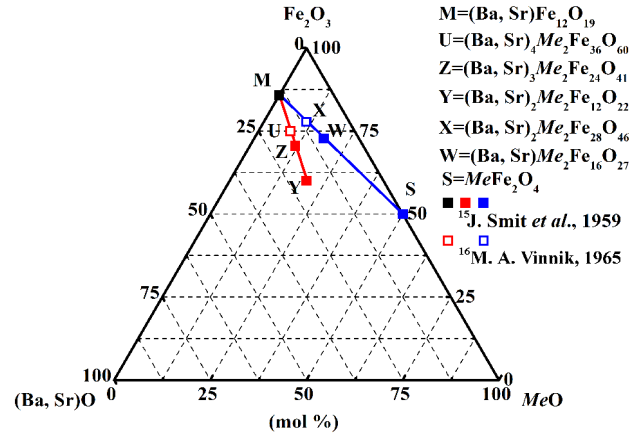

FIG. 1. The M–Y and M–S lines in the (Ba, Sr)O–$Fe_2O_3$–$Me$O ternary diagram.

## II. EXPERIMENT

Specimens of $Z_M$ and $W_M$, with the major phase of $Sr_3Co_2Z$ and $SrCo_2W$ respectively, were intentionally fabricated through a conventional solid-state reaction.[19] The stoichiometric mixtures of $SrCO_3$ (99.5$^+$ wt.%), $Co_2O_3$ (99.5$^+$ wt.%) and $Fe_2O_3$ (99.5$^+$ wt.%) were weighed, ground and calcined in air at 1250 °C for 4 h. Then the calcined powders were pulverized, ground, pressed into pellets (Φ12 ×3 mm), and sintered to ceramics in oxygen at 1150 °C for 8 h ($Z_M$) and 1250 °C for 8 h ($W_M$), respectively. Phase determination was carried out (5 ° ≤ 2θ ≤ 140 °) by X-ray diffraction (XRD) using a D8 Advance X-ray diffractometer (Bruker, Germany), with a Cu $K_α$ radiation (λ = 1.54184 Å, 40 kV, 40 mA) at a scan rate of 0.02 °/s. The magnetization ($M$) was tested in the field of 100 Oe (temperature changing from top down) by a vibrating sample magnetometer (VSM) (Changchun Yingpu Magneto-Electric Corp., China). Before the measurement of dielectricity, MD effect, ME effect and pyroelectric properties, the Ag electrodes were pasted on the polished plates (Φ10 ×1 mm) and fired in oxygen at 830 °C for 10 min. The dielectricity was determined by using a Precision Impedance Analyzer 6500B (Wayne Kerr Electronics Inc., Britain). The MD effect was measured in the field of ±10$^4$ Oe employing a homemade device. The magnetoelectric and pyroelectric properties were tested using a Physical Properties Measurement System (PPMS, Quantum Design, Inc. USA) coupled with an electrometer (Keithley 6517B, Keithley Instruments, Inc., USA). The magnetic field dependence of the electric polarization was obtained by measuring the ME current. Before this measurement, a magnetic field of 3 ×10$^4$ Oe was pre-applied. Then an electric field of 2.5 kV / cm was applied perpendicular to the magnetic field. Subsequently, the magnetic field was set to $5 \times 10^3$ Oe. After these poling procedures, the electric field was removed. The ME current was then obtained while the magnetic field was sweeping from $5 \times 10^3$ to $-3 \times 10^4$ Oe at a rate of 50 Oe / s. Before the measurement of pyroelectric current, the sample was cooled down from RT to 200 K under a poling electric field of 625 V / cm. The pyroelectric current was then obtained during the heating process (from 200 to 400 K) with a constant ramping rate of 2 K / min. For measuring the pyroelectric current under a magnetic field, the field was applied during the poling and the subsequent processes. The temperature dependence of the electric polarization was obtained by integrating the pyroelectric current.

## III. RESULTS AND DISCUSSION

### A. Phases determination

XRD patterns in Fig. 2 show that the major phase for $Z_M$ and $W_M$ is $Sr_3Co_2Z$ [PDF 19-0097, S.G.: $P6_3/mmc$ (194)] and $SrCo_2W$ [PDF 54-0106, S.G.: $P6_3/mmc$ (194)], respectively. $CoFe_2O_4$ is the main minor phase. Difficulties in synthesizing single phase of Z- / W-type hexaferrites were also stated previously.[1,6,17,18] The $Z_M$ and $W_M$ in this work represent the specimens which contain 70–90 wt.% $Sr_3Co_2Z$ phase and 70–90 wt.% $SrCo_2W$ phase, respectively. The subsequent results have high consistency and reproducibility within these phase contents.

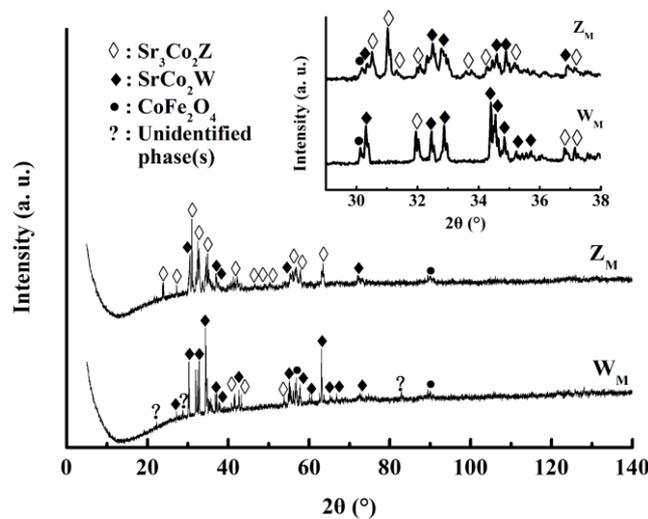

FIG. 2. XRD patterns of $Sr_3Co_2Z$ ($Z_M$) and $SrCo_2W$ ($W_M$) ceramics. The inset shows the enlargement of XRD patterns from 29 ° to 38 °.

### B. Dielectric and magnetic properties

Fig. 3 shows the temperature dependence of magnetization, dielectricity and resistivity ($M$–$T$, $\varepsilon'$–$T$, $tan\delta$–$T$ and $\rho$–$T$), and frequency dependence of dielectricity ($\varepsilon'$–$f$ and $tan\delta$–$f$ at selected temperatures) for $Z_M$ and $W_M$. Strong correlation was found between the magnetic and dielectric anomalies.

Below the Curie temperature ($T_{c,FM-PM}$ = 683 K, FM—ferrimagnetic, PM—paramagnetic),[5,18] $Ba_3Co_2Fe_{24}O_{41}$ ($Ba_3Co_2Z$) and $Ba_{1.5}Sr_{1.5}Co_2Fe_{24}O_{41}$ go through major changes in the magnetocrystalline anisotropy (MCA) with temperature increasing: Cone 1 → Plane → Cone 2 → Uniaxial.[15,18,20] However, there is no planar MCA for $Sr_3Co_2Z$: Cone 1 [52.3–58° ($\theta$), 300 K; 57°, 373 K][18,20] → Cone 2 (50.5–56°, 460–473 K)[7,18] → Uniaxial (25°, 523 K; 20°, 573 K; 16°, 623 K; 0°, 566–673 K)[7,18] ($\theta$ is the angle between $M$ and $c$-axis). The $\theta$ is less sensitive to temperature and keeps at 52.3–58° in Cone 1. But it decreases gradually to ~25° in Cone 2. The structure complexity of the $Sr_3Co_2Z$ mainly results from the $Sr^{2+}$ whose radius is comparable to that of $O^{2-}$ [$r(Sr^{2+})$ = 1.31 Å (9), 1.44 Å (12); $r(O^{2-})$ = 1.38 Å (4), the digits in brackets refer to coordination number].[21] The $Sr^{2+}$ ions prefer the oxygen positions rather than the interstitial sites, while other metal ions [$r(Fe^{3+})$ = 0.49 Å (4), 0.65 Å (6); $r(Fe^{2+})$ = 0.63 Å (4), 0.78 Å (6); $r(Co^{2+})$ = 0.58 Å (4), 0.75 Å (6)][21] are located in non-equivalent interstitial sites. $Sr_3Co_2Z$ contains Fe(Co)–O–Fe(Co) bonds in the T-blocks. Small-moment layers (S) arise in T-blocks while large ones (L) in other blocks. Since Sr ions are located near the two Fe(Co) sites, the substitution of small Sr for large Ba [*vs.* $Ba_3Co_2Z$, r ($Ba^{2+}$) = 1.47 Å (9); 1.61 Å (12)][21] increases the Fe(Co)–O–Fe(Co) bond angle ($\varphi$) ($Ba_3Co_2Z$: 116°; $Sr_3Co_2Z$: 123°) through the L / S boundary.[5] The $\varphi$ = 123° causes the magnetic frustration ($\theta$ = ~55.6°) and stabilizes a transverse-conical (T-conical) magnetic structure above RT.[7]

For the $Z_M$ ($Sr_3Co_2Z$) specimen, the $M$–$T$ curve exhibits two clear drops at around 370 K and 505 K. The later is similar with the reported while the former is different.[1,3,7,18] The anomaly at 370 K represents the onset of establishment of magnetic structure with the $P6_3/mmc$ symmetry,[7] below this temperature $Z_M$ exhibits MD / ME effects.[3,7] This means that $Z_M$ is in the ferroelectric (FE) side while not in the paraelectric (PE) side below ~370 K.[3] The anomaly at 505 K corresponds to a transition from the phase with a cone of easy magnetization into the phase where Fe and Co magnetic moments become parallel to $c$-axis.[1,3,7,18] Thus the $M$–$T$ curve can be divided into four regions. Region I, II, III and IV cover the Cone 1 (~55.6°, < 370 K), Cone 2 (~55.6° > $\theta$ > 25°, 370–505 K), Uniaxial (25° ≥ $\theta$ ≥ 0°, 505–683 K) and PM phase (> 683 K), respectively.

As to the temperature dependence of dielectricity ($\varepsilon$'—the real part of dielectric constant), a broad dielectric relaxation-peak $\varepsilon$'($T$) appears at the border of Region I and II (Fig. 3a), corresponding to the change of magnetic structure. The $tan\delta$($T$)-peak at ~550 K (Fig. 3c) coincides with the change from a cone to a uniaxial anisotropy. Temperature increasing causes $\varepsilon$'($f$)-peaks shift towards higher frequencies (Fig. 3e). The $tan\delta$($f$) decreases with frequency increasing in the whole temperature range (Fig. 3g). The high loss-factor of $tan\delta$($T$) can be attributed to the contributions from conduction loss ($10^8$ Ωcm at RT; $10^4$ Ωcm at 600 K) (Fig. 3c) and ion jump relaxation, especially at low frequencies.[22] The obtained conduction activation energy ($E_{a(c)}$) extracted from resistivity ($\rho$) is 0.72 eV, higher than that of the $Ba_3Co_2Z$ single crystal (~0.1 eV),[23] indicating the conduction carriers in the $Z_M$ are from the second ionization of oxygen vacancies.[24] In the polycrystalline ceramic of $Sr_3Co_2Z$ hexaferrite, medium resistance (*vs.* $Ba_3Co_2Z$, low resistance) grains are separated by highly resistive grain-boundaries. The preparation in oxygen can decrease $Fe^{2+}$ concentration and then reduce the hopping interchange of electrons between $Fe^{2+}$ and $Fe^{3+}$ ions. Thus the oxygen sintering favors a high resistivity.[1]

Below the Curie temperature ($T_{c,FM-PM}$ = 763 K; 728 K)[5,25], $BaCo_2Fe_{16}O_{27}$ ($BaCo_2W$) goes through the following major changes in MCA with temperature increasing: Cone 1 (~70°, 2–453 K; 68.5–70°, 300 K)[5,25,26] → Cone 2 (70° > θ > 0°, 453–553 K)[25] → Uniaxial (0°, 553–763 K)[25]. For the $W_M$ specimen ($SrCo_2W$), the *M–T* curve exhibits one drop at ~470 K (Fig. 3b). When temperature increases, the *M* decreases steeply at 470–520 K. Thus the *M–T* curve can also be divided into four regions. Region i, ii, iii and iv cover the Cone 1 (< 470 K), Cone 2 (470–520 K), Uniaxial (520–763 K) and PM phase (> 763 K), respectively. The difference is obvious between the two specimens. The $\theta$ decreases quickly from 70° to 0° within 100 K (Region ii)[25] and then keeps at 0° till $T_{c,FM-PM}$ (Region iii) in the $W_M$, but the $\theta$ decreases gradually from ~55.6° to 0° spanning the Region II and III in the $Z_M$. Compared with the $Z_M$, there is a similar dielectric anomaly of $\varepsilon'(T)$ in the $W_M$ at the border of Region i and ii. Other characteristics for the $W_M$ are analogous with those of the $Z_M$ (Fig. 3d, 3f and 3h).

The variation of $\theta$ as a function of temperature is summarized for Y-, Z- and W-type hexaferrites in Fig. 4.[4,7,15,18,20,25,27] Below the critical temperature, Y- and Z-type hexaferrites exhibit ME effect. It is amazing for us to find that there exists a golden ratio point (critical $\theta$, 55.6° ≈ 0.618 × 90°) which is correlative to MD / ME effects for $Sr_3Co_2Z$ below 400 K (Fig. 4).[4,7,15,18,20,25,27] Besides, for $Ba_2Co_2Y$ and $Ba_3Co_2Z$, which exibit ME effect, their $\theta$s are also approaching to the golden ratio point (55.6°) below critical temperature. There may be some important physical mechanisms accounting for this phenomenon which need further investigation.

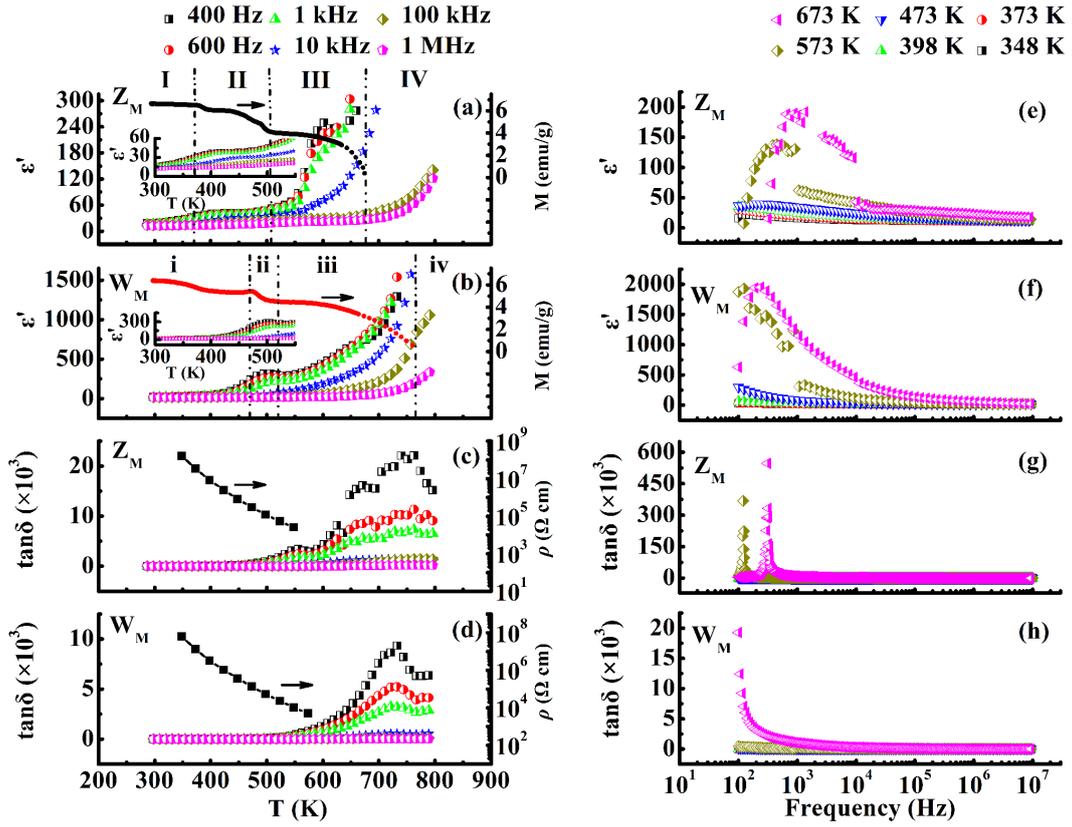

FIG. 3. Temperature dependence of magnetization, dielectricity and resistivity, and frequency dependence of dielectricity for $Sr_3Co_2Z$ ($Z_M$) and $SrCo_2W$ ($W_M$) specimens. (a) $M$–$T$ and $\varepsilon'$–$T$ for $Z_M$ (the inset is the enlargement at 300–550 K); (b) $M$–$T$ and $\varepsilon'$–$T$ for $W_M$ (ibid); (c) $tan\delta$–$T$ and $\rho$–$T$ for $Z_M$; (d) $tan\delta$–$T$ and $\rho$–$T$ for $W_M$; (e) $\varepsilon'$–$f$ for $Z_M$; (f) $\varepsilon'$–$f$ for $W_M$; (g) $tan\delta$–$f$ for $Z_M$; (h) $tan\delta$–$f$ for $W_M$.

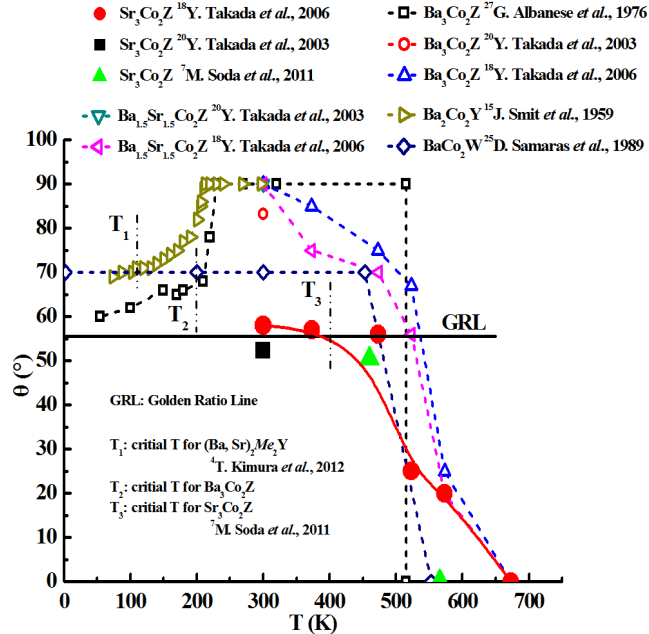

FIG. 4. The golden ratio line (GRL) of the magnetocrystalline anisotropy (MCA) transition in hexaferrites.

So the magnetic and dielectric anomalies are all associated with the MCA transitions in both specimens ($Z_M$: ~370 K; $W_M$: ~470 K). The $Z_M$ ($Sr_3Co_2Z$) exhibits MD / ME effects above RT but $W_M$ ($SrCo_2W$) does not in any temperature range.[4,5,7] That means the magnetic and dielectric anomalies are characteristic properties (*e.g.*

$Z_M$ at ~370 K) but do not satisfy a sufficient condition for the MD and ME coupling.

## C. Magnetodielectric properties

Fig. 5 shows the magnetic-field dependence of dielectricity, *i.e.*, the MD effect (measured at 100 kHz at RT). The $Sr_3Co_2Z$ ($Z_M$) has a smaller $\varepsilon'$ than that of the $SrCo_2W$ ($W_M$) in the whole magnetic field ($H = -10^4$ Oe ~ $10^4$ Oe) (Fig. 5a), while the *tanδ* of the $Z_M$ is larger than that of the $W_M$ (Fig. 5b). For the $W_M$, its $\varepsilon'$ and *tanδ* are not affected by the $H$ up to $10^4$ Oe. For the $Z_M$, the $\varepsilon'$ shows magnetic-field dependence. It decreases with $H$ increasing and is field-direction dependent. When the $H$ increases from 0 Oe to $10^4$ Oe, the $\varepsilon'$ shows a maximum (~16.36) at 0 Oe and decreases gradually to 16.31 at $7 \times 10^3$ Oe, then becomes nearly constant above $7 \times 10^3$ Oe. The *tanδ* is almost magnetic-field independent, similar with the previously reported.[1] Accordingly, the dielectric-constant-change ratio $[\Delta\varepsilon'/\varepsilon'(0) = (\varepsilon'(H) - \varepsilon'(0))/\varepsilon'(0)]$ increases gradually with $H$ increasing from 0 to ~$7 \times 10^3$ Oe, and then becomes nearly constant (–0.3%) above $7 \times 10^3$ Oe (Fig. 5c). This ratio is comparable with those obtained by Kitagawa *et al*. (–3%, 100 kHZ)[1] and Zhang *et al*. (–4%, 50 MHz)[3]. It has been proposed that $\Delta\varepsilon'$ is proportional to the square of $M$ based on the framework of Ginzburg–Landau theory, *i.e.*, $\Delta\varepsilon' \propto \gamma M^2$, where $\gamma$ is a coupling constant and is negative for $Sr_3Co_2Z$.[3,28] In the $Z_M$ of this work, the $M$ increases with $H$ increasing, leading to the $\varepsilon'$ decreasing. The $\varepsilon'$ shows a maximum in the vicinity of PE–FE transition induced by $H$.[29] When $M$ is saturated at ~$7 \times 10^3$ Oe, the $\varepsilon'$ becomes a constant. The $Z_M$ presents a negative MD effect at RT, because it is in the FE side.[3] Therefore the MD effect in the $Z_M$ corresponds to the fact that the PE–FE transition is induced by $H$ and the $\varepsilon'$ decreases with $H$.

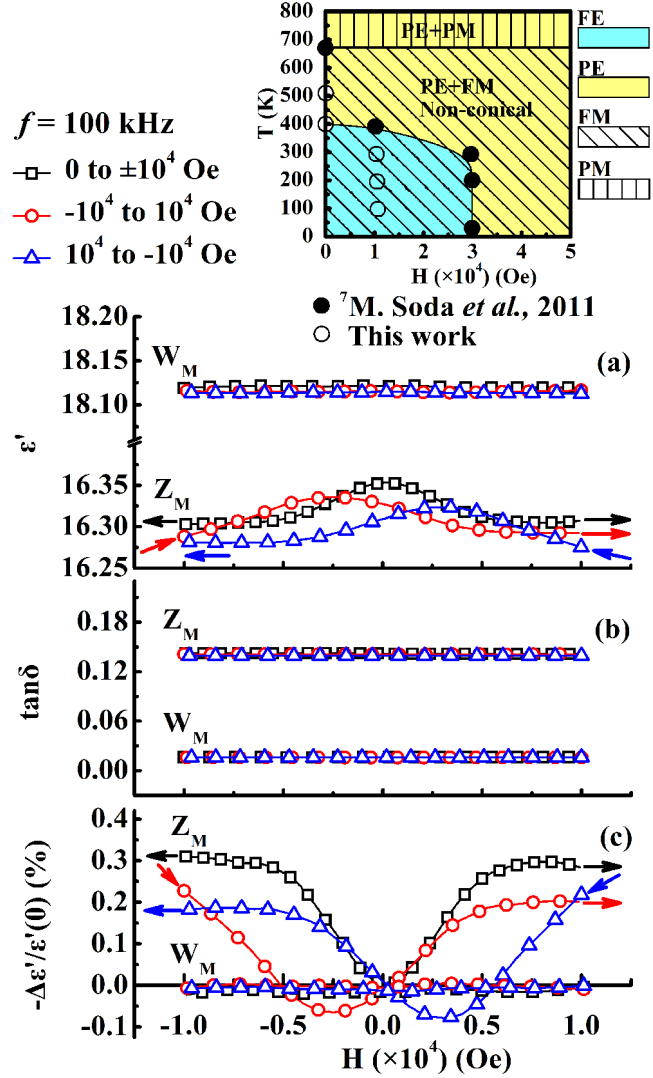

FIG. 5. Magnetic-field dependence of dielectric constant (a), loss factor (b) and dielectric-constant-change ratio (c) for $Sr_3Co_2Z$ ($Z_M$) and $SrCo_2W$ ($W_M$) specimens (100 kHz; RT). The inset is magnetic phase diagram for $Sr_3Co_2Z$ with data from the reference (solid black dots)[7] and this work (hollow dots).

## D. Magnetoelectric properties

Fig. 6 shows the magnetic field dependence of electric polarization and magnetoelectric current (in the inset) for $Sr_3Co_2Z$ ($Z_M$). The magnetoelectric current for $Z_M$ shows remarkable dip and peak structure centered at around $-500$ and $-8 \times 10^3$ Oe at 100, 200 and 300 K. The magnetic field dependence of the electric polarization, calculated by integrating the megnetoelectric current, reveals the ME coupling. There is almost no spontaneous polarization at zero magnetic field. By applying a increasing magnetic field, the electric polarization increases rapidly and reaches a maximum at about $-4 \times 10^3$ Oe. It decreases then and vanishes at around $-1.3 \times 10^4$ Oe where the system becomes a simple ferrimagnet.

The results are some similar with those from Kitagawa et al.[1] and Soda et al..[7] Differences are also in existence. The electric polarization decreases with temperature increasing in Kitagawa et al.'s work. But the electric polarization at 200 K is larger than that at 100 K in this work. The same trend was observed by Soda et *al.*, in which the polarization of $Sr_3Co_2Z$ at 200 K is larger than that at 10 K.[7] Besides, the magnitude of electric polarization in this work is smaller than those reported previously. The difference could be attributed to the different synthesis process of $Sr_3Co_2Z$.[7]

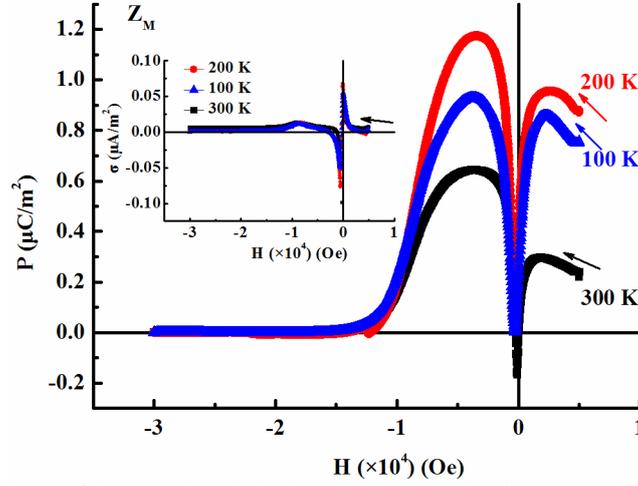

FIG.6. Magnetic field dependence of the electric polarization and magnetoelectric current (in the inset) for $Sr_3Co_2Z$ ($Z_M$) at different temperatures.

### E. Pyroelectric properties

Fig. 7 shows the pyroelectric current and electric polarization (in the inset) as a function of temperature for $Sr_3Co_2Z$ ($Z_M$) under different magnetic field. There is no pyroelectric current found below 300 K during 0 to $1 \times 10^4$ Oe. With temperature increasing, the pyroelectric current increases gradually, reaches its maximum at ~380 K and then decreases. The peak-temperature corresponds well to that where the magnetic and dielectric anomalies present. The intensity of the pyroelectric current increases with magnetic field increasing, leading to an increased electric polarization (the inset of Fig. 7). This indicates that the electric polarization is enhanced by the electric field.

The pyroelectric current peak centered at 380 K implies an intrinsic and over-room-temperature ferroelectricity for $Z_M$, even when the magnetic field is zero. For comparison, the ferroelectricity of $Sr_3Co_2Z$ at zero magnetic field was claimed by Zhang *et al.*[3] and Wu *et al.*[10] but there was no direct evidence provided. The magnetic-field-enhanced electric-polarization has been reported in $DyMnO_3$ thin films which is understood by the Dy-Mn spin interaction.[30] In the case of $Sr_3Co_2Z$, its electric polarization originates from the T-conical structure through the inverse Dzyaloshinskii-Moriya (DM) mechanism. The spin structure of $Sr_3Co_2Z$ could be slightly modified by the proper magnetic field, resulting in the increase of electric polarization.

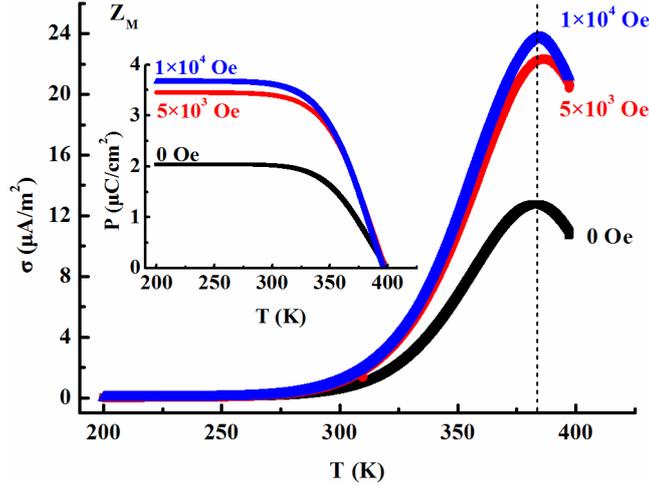

FIG. 7. Temperature dependence of pyroelectric current and electric polarization (in the inset) for $Sr_3Co_2Z$ ($Z_M$) under different magnetic field.

## F. Discussion

Because of its complex T-conical structure which evolves with temperature, the dielectric response in the $Sr_3Co_2Z$ ($Z_M$) includes several contributions. Each of them is related to a particular microscopic mechanism and becomes significant in a particular frequency ($\omega = 2\pi f$) / temperature interval. The $\varepsilon'$ can be represented as:[31]

$$\varepsilon'(\omega,T,M) = 1 + \sum_i \chi_i(\omega,T,M) = 1 + \chi_\infty(\omega,T) + \chi_{FE}(\omega,T,M) + \chi_M(\omega,T,M) + \chi_{LF}(\omega,T) \qquad (1)$$
$$= \varepsilon'_\infty(\omega,T) + \chi_{FE}(\omega,T,M) + \chi_M(\omega,T,M) + \chi_{LF}(\omega,T)$$

where $\varepsilon'_\infty \propto N_D$ ($N_D$ is dipole density in material) is related to the polarization at high frequencies, $\chi_{FE} \propto \omega^{-2} \propto (T-T_{c,FE-PE})^{-1}$ coincides with the contribution of the FE order,[31] $\Delta\chi_M \propto \gamma M^2$ ($\gamma < 0$) is derived from the framework of Ginzburg–Landau theory as mentioned before, $\chi_{LF}$ is the low-frequency susceptibility (where the conductivity contribution $\chi_\sigma \propto T/\omega$ is concluded)[32] due to the relaxation of domain walls in polydomain crystals, mobile charge carriers, and crystal defects, *etc.*[31].

Because $Sr_3Co_2Z$ ($Z_M$) locates in its FE zone (Region I) below 370 K, $\chi_{FE}$ is dominant at low frequencies (400 Hz - 1 kHz) (Fig. 3 and Fig. 5). A dielectric anomaly appears at 370 K which coincides to the magnetic structure change. The dielectric anomalies become weaker with frequencies increasing (10 kHz - 1 MHz) because the $\omega$ is more sensitive to the $\chi_{FE}$ than the $T$ or $M$. When a magnetic field ($H$) exists in Region I (below 370 K), $\chi_M$ brings a supplementary contribution to $\varepsilon'$ before $M$ reaches saturation.

For a Z-type hexaferrite, the room-temperature MD effect is attributed to the change of a T-conical spin structure and spin-phonon coupling.[3] The room-temperature ME effect is understood in terms of the appearance of electric polarization which is induced also by a T-conical spin structure through the inverse DM interaction.[7] The above T-conical spin structure refers to the antiphase arrangement of the magnetic moments between neighboring T-blocks, which exists in $Sr_3Co_2Z$ ($Z_M$) but absent in $SrCo_2W$ ($W_M$).[1,4] As mentioned before, the Fe(Co)–O–Fe(Co) bond angle ($\varphi$) in $Sr_3Co_2Z$ arouses the magnetic frustration, stabilizes a noncollinear magnetic structure above RT and contributes to the ME performance.[7] Thus the $Sr_3Co_2Z$ has a ME-response driven by phase competition (*i.e.*, spin-driven ferroelectrics).[5]

## IV. CONCLUSIONS

In summary, dielectric anomalies near the magnetic phase transition temperatures are reported firstly for Z-type hexaferrite $Sr_3Co_2Fe_{24}O_{41}$ ($Sr_3Co_2Z$) at ~370 K and $SrCo_2Fe_{16}O_{27}$ ($SrCo_2W$) at ~470 K in this work. Correlation between the anomalies was investigated considering the crystal and magnetic structures. It is concluded that magnetocrystalline anisotropy (MCA) transition induces the anomalies, which are characteristic properties but do not satisfy a sufficient condition for the magnetodielectric / magnetoelectric coupling. T-block crystal structure that exists in $Sr_3Co_2Z$ but absent in $SrCo_2W$ is proposed to contribute to the observed results. The composition with full Sr substitution for Z-type hexaferrites and high resistivity from oxygen sintering cause the magnetic frustration and stabilize a transverse-conical (T-conical) magnetic structure.

An over-room-temperature ferroelectric effect is conformed directly and reported firstly in $Sr_3Co_2Z$ by the pyroelectric current measurements, where the peak-temperature (~380 K) is corresponding well to that where the magnetic and dielectric anomalies present. This work helps understanding the origin of MD / ME coupling effects at a low magnetic field and above room temperature.

## ACKNOWLEDGEMENTS

The authors gratefully acknowledge Prof. Z.-G. Sun and Prof. J.-F. Wang in WUT, Prof. J. Xu in WIT, Prof. Z.-C. Xia, G.-F. Fan and Z.-M. Tian in HUST, and Dr. J. Lu in IPCAS. This work was supported by grants from State Key Laboratory of Advanced Technology for Materials Synthesis and Processing (WUT, China) (Grant Nos. 2010-PY-4, 2013-KF-2, 2014-KF-6), the Open Research Foundation of Key Laboratory of Nondestructive Testing (NHU, China) (Grant No. Zd201329002), the National Natural Science Foundation of China (Grant No. 51172049) and the Special Prophase Project on National Basic Research Program of China (Grant No. 2012CB722804).